# Readout electronics of silicon detectors used in space cosmic-ray charges measurement

ZHANG Fei (张飞)[1] FAN Rui-Rui (樊瑞睿)[1] PENG Wen-Xi (彭文溪)[1]
DONG Yi-Fan (董亦凡)[1,2] GONG Ke (龚轲)[1] LIANG Xiao-Hua (梁晓华)[1]
Liu Ya-Qing (刘雅清)[1,2] WANG Huan-Yu (王焕玉)[1]

[1]Institute of High Energy Physics, Chinese Academy of Sciences, Beijing 100049, China
[2]University of Chinese Academy of Sciences, Beijing 100049, China

**Abstract:** A readout electronics used in space cosmic-ray charges measurement for multi-channel silicon detector and its performance test results are introduced in this paper. A 64-channel charge sensitive ASIC (VA140) from IDEAS company is adopt in this method. With its features of low power consumption, low noise, large dynamic range and high integration, it can be used in future particle detecting experiments base on silicon detector.
**Key words:** VA140, ASIC, readout electronics, silicon detector, charge measurement

## 1. Introduction

With the development of semiconductor technology, the fully-depleted silicon detector is widely used in particle detection physics, medical instruments and astrophysics for its features of low noise, high counting rate and good linearity. In space astrophysics detection area, it is also used for measuring the charges of ionizing particles in many projects, such as the Payload for Antimatter Matter Exploration and Light-nuclei Astrophysics (PAMELA, 1997) [1], the Advanced Thin Ionization Calorimeter (ATIC, 1999) [2], and the Alpha Magnetic Spectrometer 02 (AMS-02, 2011) [3].

The energy lost by a heavy, charged particle moving through a given thickness of fully-depleted silicon detector is due to ionization. The mean energy loss is given by the Bethe-Bloch formula :

$$-\frac{dE}{dx} = \frac{4\pi}{m_e c^2} \cdot \frac{nz^2}{\beta^2} \cdot \left(\frac{e^2}{4\pi\varepsilon_0}\right)^2 \cdot \left[\ln\left(\frac{2m_e c^2\beta^2}{I\cdot(1-\beta^2)}\right) - \beta^2\right]$$

Where $\varepsilon_0$, $I$, $x$, $e$, $m_e$, $\beta$ and c are, respectively, the vacuum permittivity, mean excitation potential of the target, distance travelled by the particle, charge of the electron, rest mass of the electron, particle velocity, and speed of light. $E$ is the energy of the particle and $Z$ is the particle charge. From the Bethe-Bloch formula, the mean energy loss is proportional to $Z^2$. The accurate dE/dx measurements can be used in cosmic nuclei identification.

A typical readout system for charges measurement consists of charge sensitive pre-amplifier, filter-shaper, peak holding circuit, analog-to-digital converter and controlling circuit. The large area silicon detectors used for space cosmic-ray charges measurement, such as silicon matrix detector, silicon pixel detector, silicon micro-strip detector, often contains thousands of channels. A great challenge will be brought by the readout electronics for large number of readout channels in space mission. It would be complex in integration, reliability and power dissipation if traditional readout way with discrete transistors, diodes and amplifiers is adopted. Since the power, weight and volume are very limited onboard the satellite platform, the application specific integrated circuits (ASIC) with features of low noise, low power dissipation and high integration are significant in the readout electronics for multi-channel silicon detectors in space exploration area [4].

## 2. Overview of VA140

VA140 is a 64-channel, low noise and high dynamic range charge measurement ASIC designed by IDEADS (Norway). Its internal architecture is shown in figure 1. Each channel contains a charge sensitive preamplifier, a shaper circuit and a sample / hold circuit. An analog multiplexer controlled by the shift register in the left side are adopted to transfer the holding signals of 64 channels to the differential output port. Besides, there are calibration facilities by means of which the linearity of each channel can be calibrated by external injected charges [5].

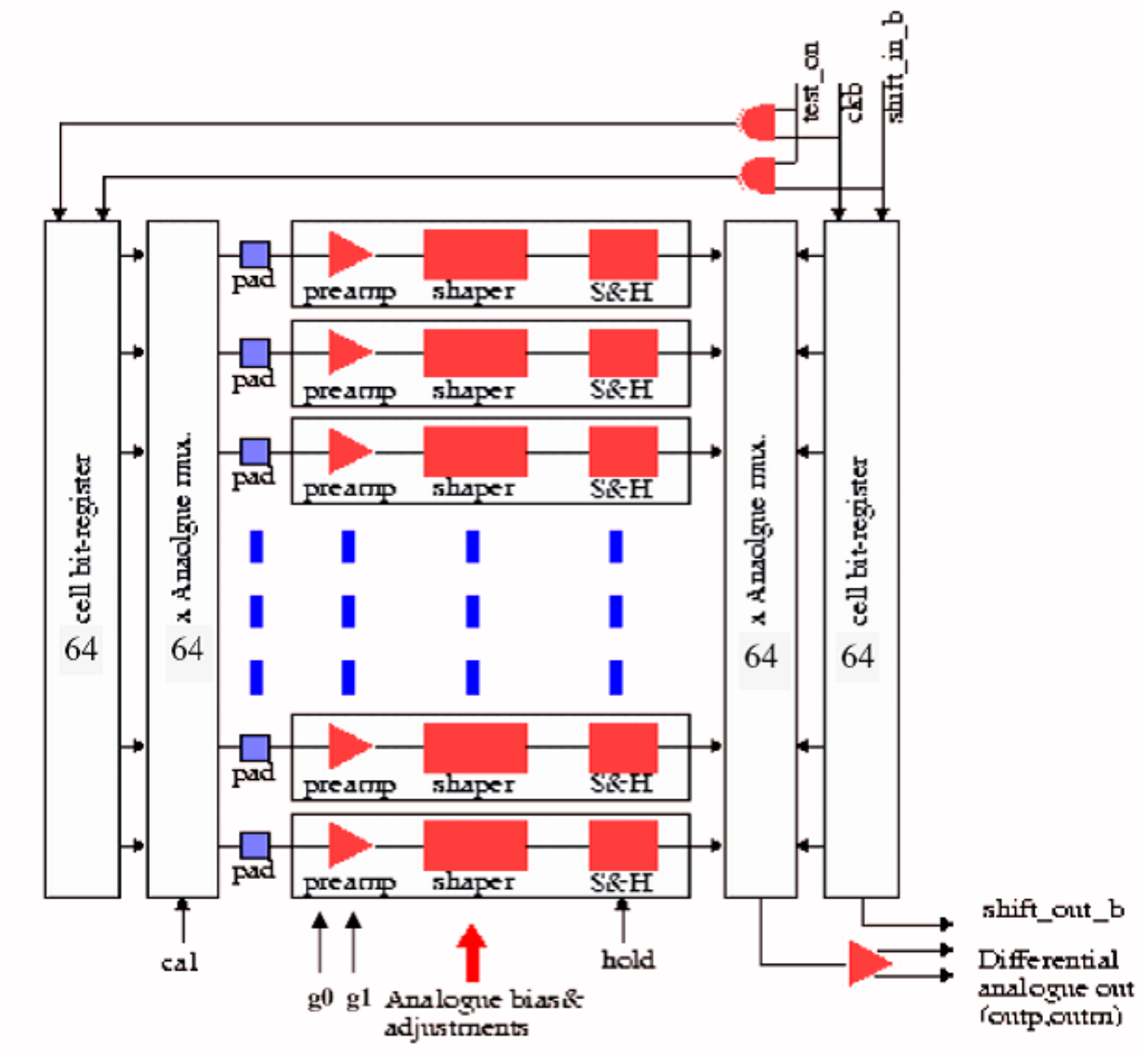

Fig. 1 VA140 internal architecture

## 3. Development of readout electronics

### 3.1 Hardware for readout system

As shown in Fig.2, the readout electronics system of the multi-channel silicon detectors is comprised of a detector board, an ASIC board, a control board and a PC. A shielded box is adopted to prevent the detectors and ASIC from being interfered by the surrounding light and EMI. The control board is responsible for controlling VA140 work, digitalizing the analog signal and transferring data packets between the board and the PC through a USB2.0 bus. It contains a power generator module (HV and regulator), a controlling FPGA (XC3S500E), a 14-bit 3MSPS ADC (AD9243) and a USB interface chip (CY7C68013). A level shift circuit due to the mismatched level voltage between FPGA and VA140 is used in the control board to convert the FPGA driver signals from 0~3.3V to -2V~+1.5V. In order to fully deplete the silicon detectors, a HV module (E10222 from EMCO company) is adopted to generate the 80V bias voltage.

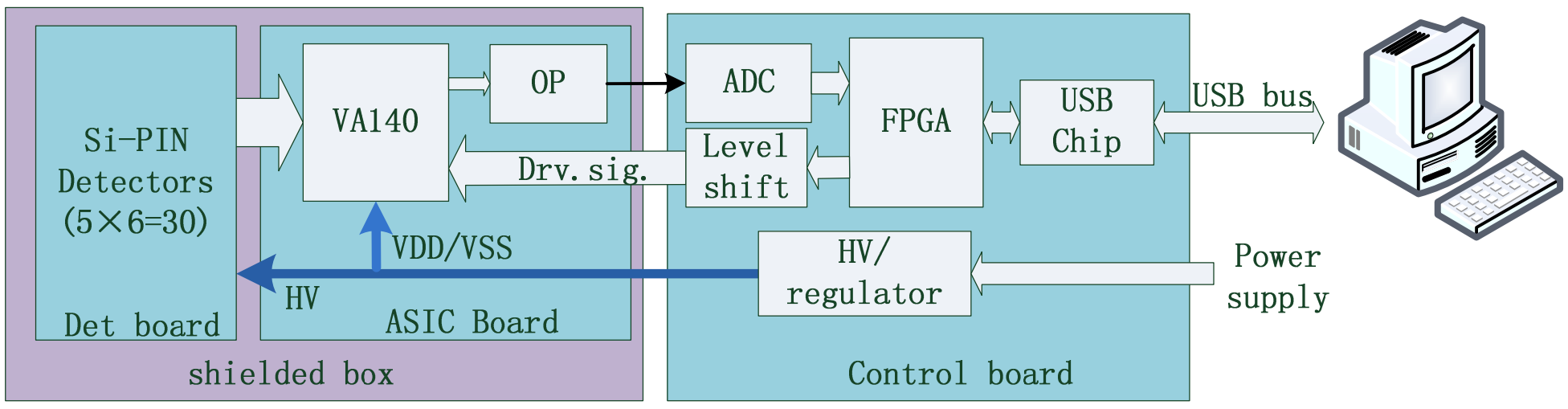

Fig. 2 Readout electronics block diagram

There are 30 Si-PIN detectors mounted in the detector board. The area and thickness of each detector are $5cm^2$ and 300 um, respectively. The detectors are connected to the VA140 channels (ch-1 to ch-30) separately through a flexible cable. The front electronics for each Si-PIN detector is shown in Fig.3. Two 10MΩ resistors and two 10 nF capacitors near the detector are adopted to act as a filter for the HV bias. When a charged particle or a high-energy photon goes into the fully-depleted area of the detector, the electrons and holes will be generated and moved to N side and P side, respectively. A 100MΩ resistor is used to sample the positive signal in the P side, and the charges will be injected to the preamplifier of a certain VA140 channel through a 1 nF coupling capacitor.

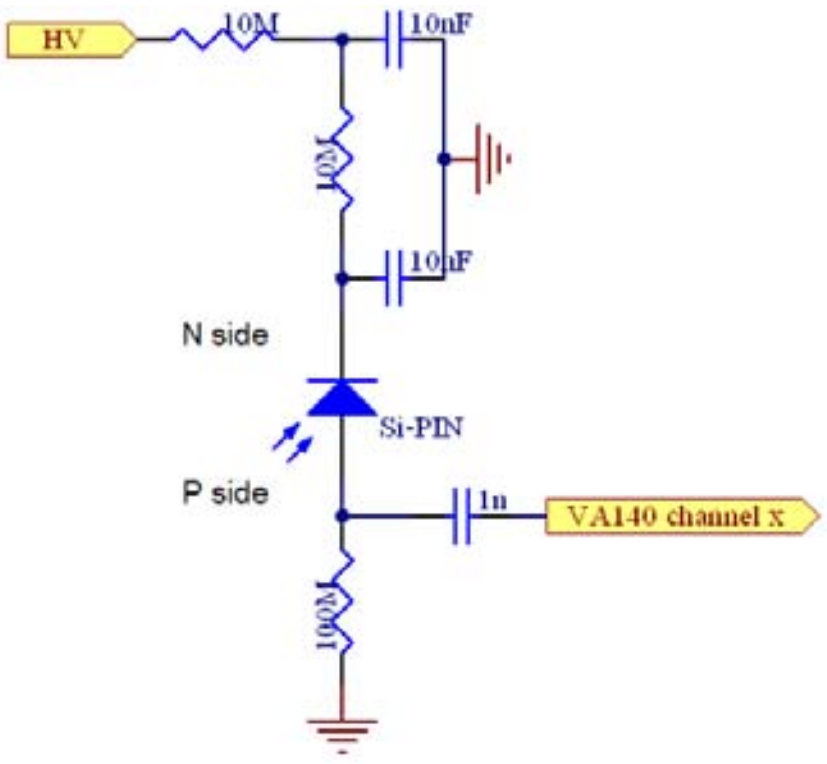

Fig. 3 Front electronics of Si-PIN detector

The all readout system is powered by external +5V and -5V DC modules whose currents are 130mA and 40mA, respectively. The currents of -2V and +1.5V for VA140 generated from -5V and +5V through low-dropout voltage regulators, are about 8mA and 2mA, respectively. The power dissipation for each channel is about 0.3mW.

### 3.2 Software for readout system

The software for the readout system includes a logic program in the FPGA and an application in the PC.

The FPGA logic is designed by the High Speed Integrated Circuit Hardware Description Language (VHDL) and synthesized by the Xilinx ISE tools. During normal work, the FPGA receives the configuration commands from the PC through USB bus firstly, then drives the VA140 and ADC working, and transfers the data to the PC. The readout timing of VA140 is shown in Fig.4. When an event happens, the FPGA will give the [holdb] signal after fixed 6.5us to hold the shaper signals of all the VA140 channels. Then the FPGA will give the [shift_inb] and [ckb] signals to control VA140 to shift out the sampling analog signals channel by channel. The analog signal outputed from VA140 common path will be digitalized by ADC at the meantime. The shift clock [ckb] is 1Mhz, so it will take 64 us to read all the 64 channels for one event.

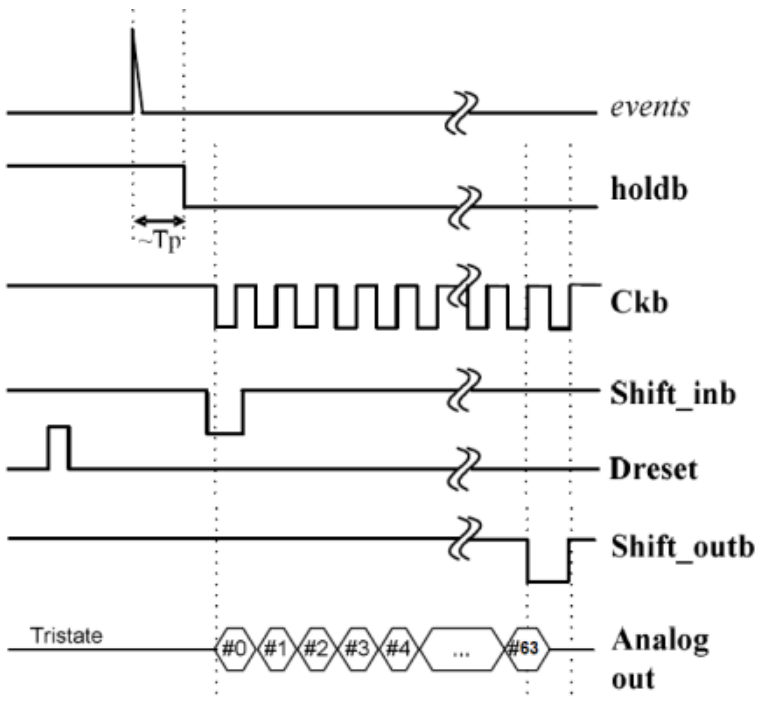

Fig. 4 Readout Timing of VA140

The application in the PC is developed by C language based on Labwindows/CVI, as shown in Fig.5. It provides a user-friendly Graphical User Interface (GUI) by means of which we can configure the parameter of the readout system, read the data packet through USB bus, process and display online.

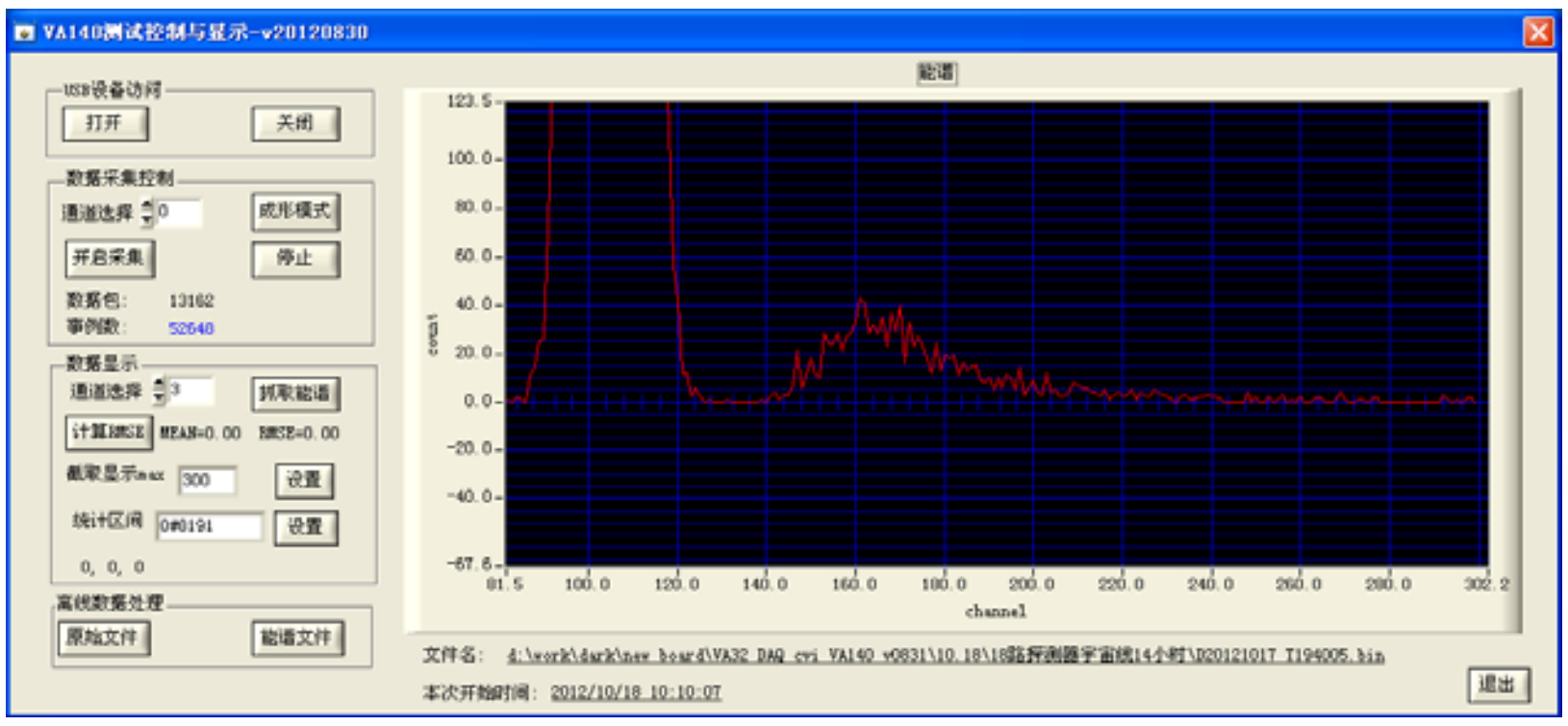

Fig. 5 PC software based on Labwindows/CVI

## 4. performance of the readout system

### 4.1 linearity test for VA140

The VA140 charges injection from silicon detectors can be simulated by a pulse voltage going through a capacitance. The value is equal to the result of multiplying the pulse amplitude by the coupling capacitance. In order to measure the linearity of the VA140, a serial of pulses with increasing amplitudes from 10mV to 200mV are generated by DG4162 signal generator. A 2pF coupling capacitance is connected to the ch-1 of VA140. The linearity curve of VA140 is shown in Fig.6, the INL is better than +/-1.5% in 0 to 200fC range.

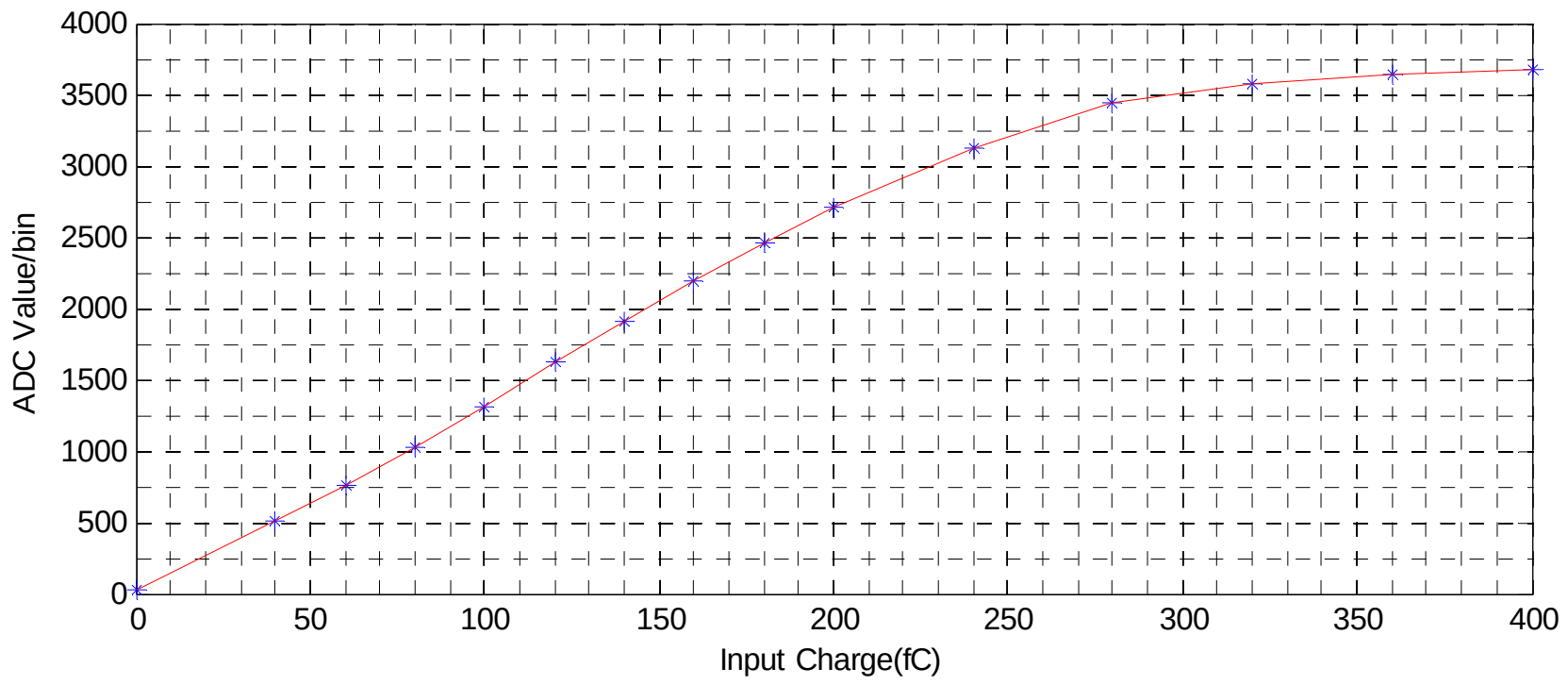

Fig.6 linearity curve of VA140

### 4.2 Noise slope test with different input capacitors

Various capacitors (0pF to 510pF) were added between GND and the input pad of ch-1. The noise slope is shown in Fig.7, approximately 0.0018fC/pF (i.e. 11e/pF).

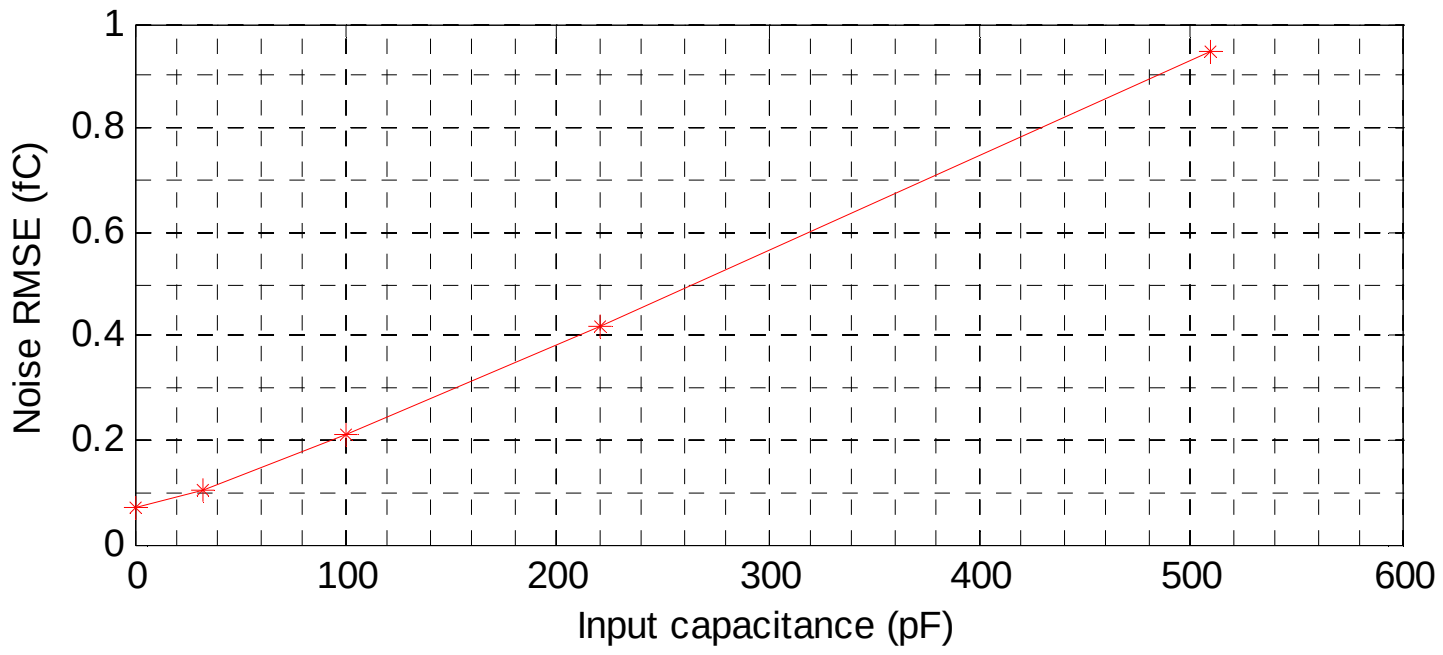

Fig.7 Noise slop curve of VA140

### 4.3 Pedestal and noise test

The VA140 channels of ch-1 to ch-30 are connected to the Si-PIN detectors, and the left are not connected. The pedestal and noise are tested under 100Hz random trig. As shown in Fig.8, the noise values (RMSE) of the channels with detectors connecting and without detectors connecting are about 0.35fC and 0.07fC, respectively. The increasing noise with detectors connecting is due to the capacitance and leak current of the detector.

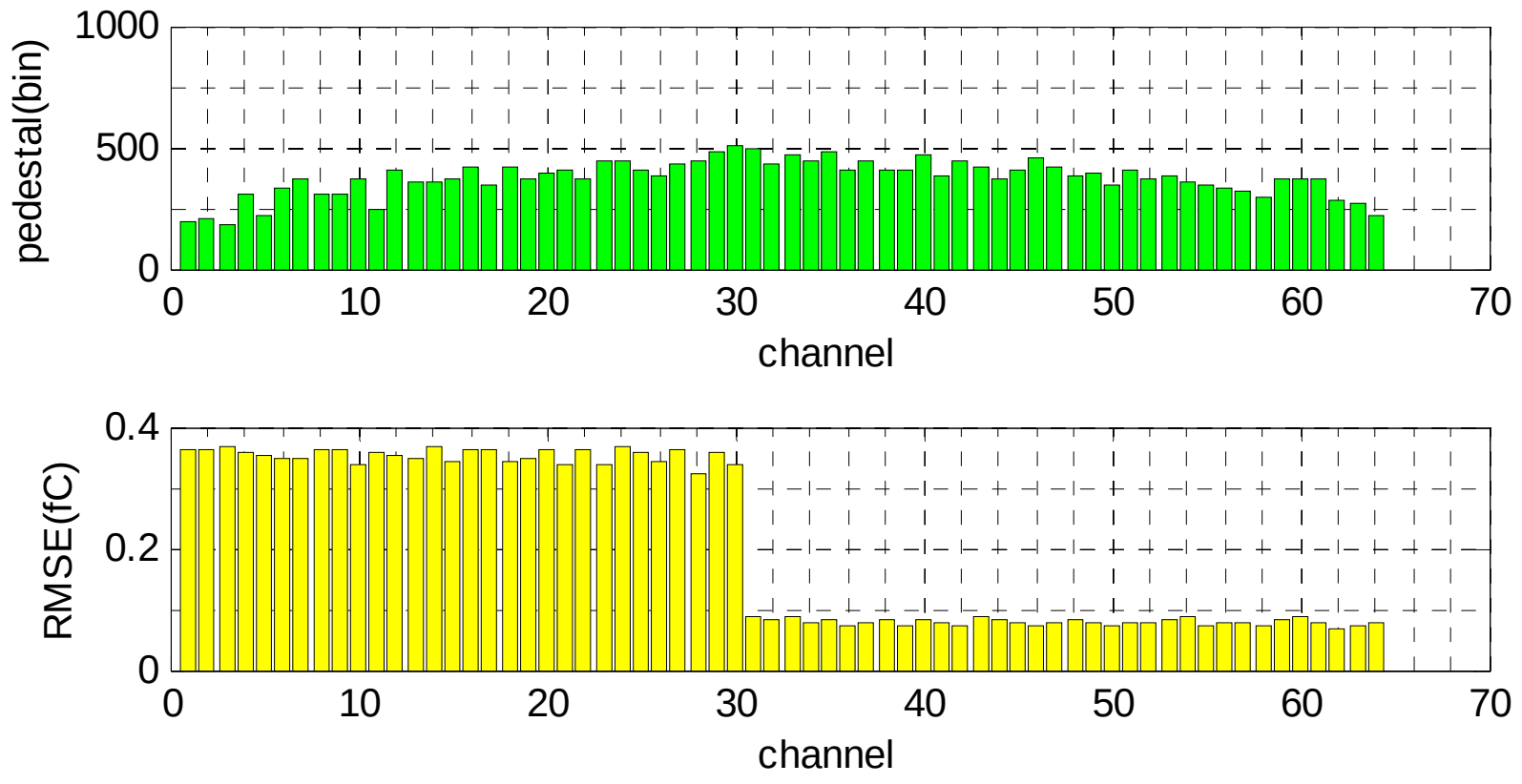

Fig.8 pedestal and noise test of VA140

### 4.4 Cosmic ray test

The cosmic ray (Muon) test system in laboratory is shown in Fig.9. Two scintillators on and under the silicon detectors are used to generate the coincidence trig when a Muon particle goes through the detectors. The control board will read all channels of VA140 and send the data to the PC after every trig.

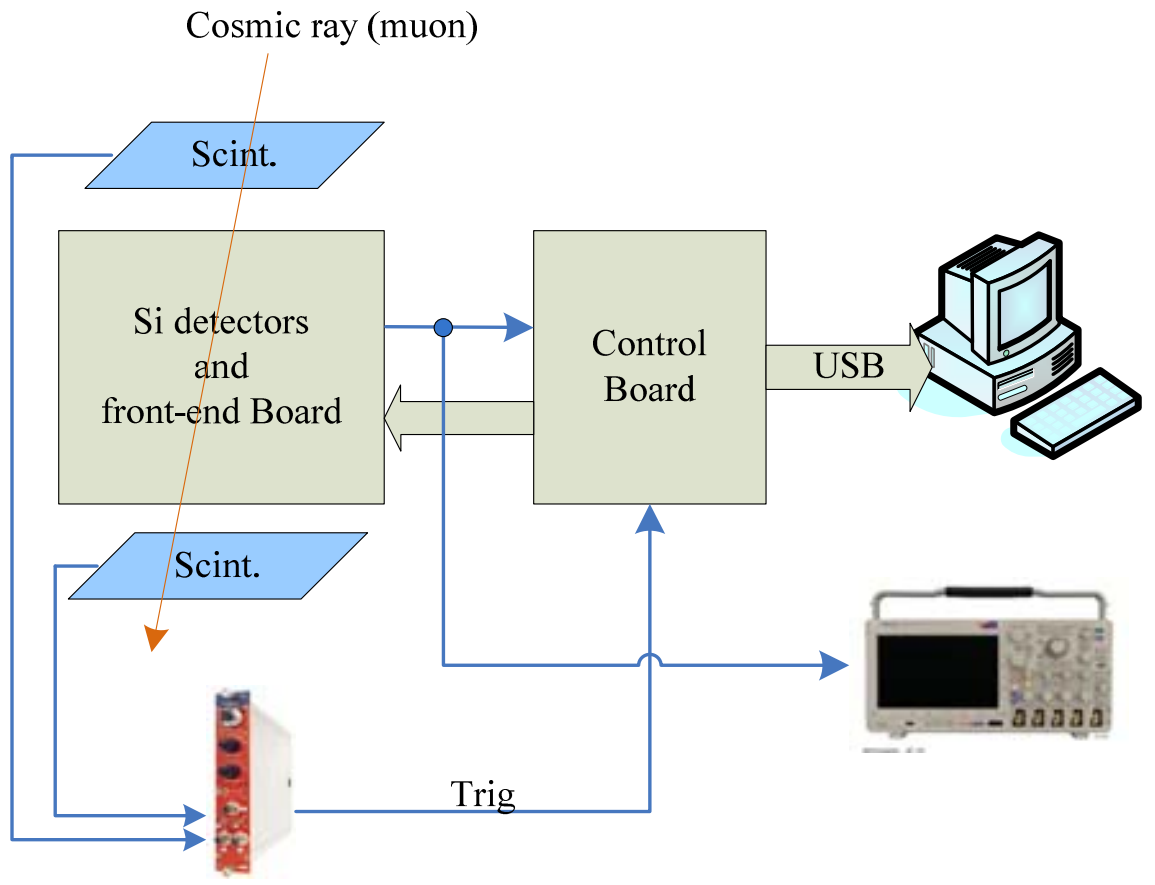

Fig. 9 Cosmic ray test system in laboratory

The cosmic ray test system work stably in successive 48 hours test. The accumulative energy spectrum is shown in Fig.9. The MIP spectrum of Muon obeys Landau distribution, and can be obviously distinguished from the pedestal.

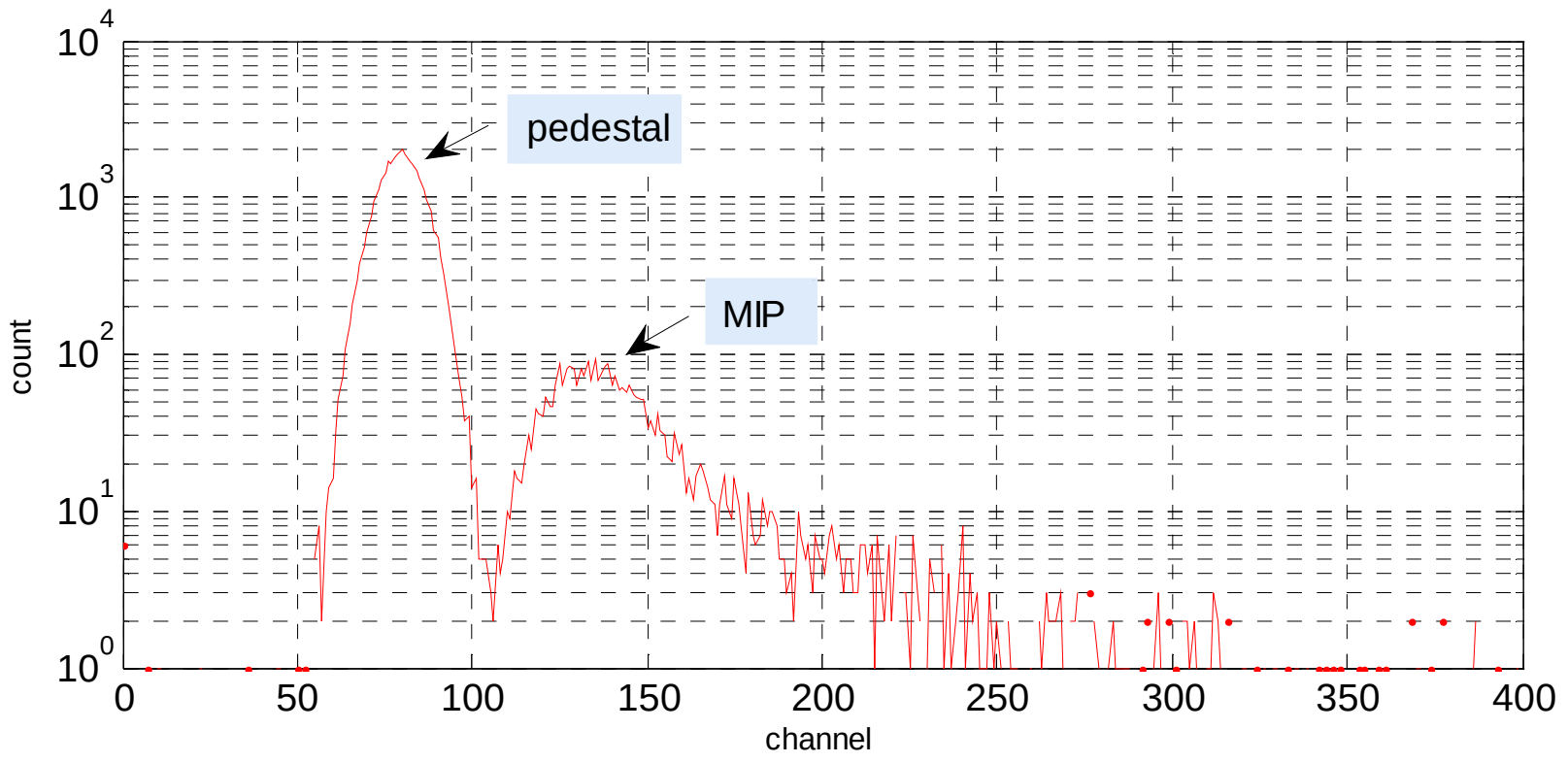

Fig. 10 MIP curve of cosmic ray (muon)

## 5. Conclusion

A readout electronics system for silicon detectors to measure the space cosmic-ray charges are presented in this paper. It shows that the readout system has features of low noise (less than 0.1fC when silicon the detector connects), low power dissipation (about 0.3mW/channel), high dynamic range (200fC) and high integration (64 channels in one chip). It can be adopted in many space particle detection experiments.

In order to get higher dynamic range for the readout system, some works still need to be done. The thickness of the detector can be decreased to reduce the energy deposition, and the value of the preamplifier feedback capacitor can be increased to expand the dynamic range of the ASIC.